\documentclass[12pt,a4paper]{article}
\usepackage{graphicx}

\newenvironment{inmargins}[1]{
    \begin{list}{}{
        \leftmargin=#1
        \rightmargin=#1
        \parsep=0pt
        \partopsep=0pt
    }
    \item[]
}{\end{list}}

\newcommand{\be}{\begin{equation}}
\newcommand{\ee}{\end{equation}}
\newcommand{\bea}{\begin{eqnarray}}
\newcommand{\eea}{\end{eqnarray}}

\makeatletter
\def\section{\@startsection{section}{1}{\z@}{-3.5ex plus -1ex minus -.2ex}{2.3ex plus .2ex}{\normalsize \bf}}
\def\subsection{\@startsection{subsection}{2}{\z@}{-3.25ex plus -1ex minus -.2ex}{1.5ex plus .2ex}{\normalsize \sl}}
\def\subsubsection{\@startsection{subsubsection}{3}{\z@}{-3.25ex plus -1ex minus -.2ex}{1.5ex plus .2ex}{\normalsize}}
\makeatother

\makeatletter
\def\@evenhead{
    \vbox{\hbox to\hsize{\bf \thepage \hfill \sl \evenmark}
    }
}
\def\@oddhead{
    \vbox{
        \hbox to\hsize{\oddmarkA \oddmarkB \hfill \oddmarkC}
        \hbox to\hsize{\hfill \oddmarkD}
        \vspace{.03in}
        \hbox to\hsize{\hfill \oddmarkE}
        \vspace{.03in}
        \hbox to\hsize{\hfill \oddmarkF}
    }
}
\def\@evenfoot{\vbox{\hbox to\hsize{\hrulefill}
    \vspace{-40pt} \hbox to \hsize{\today \sl \footmsg \hfill}
    }
}
\def\@oddfoot{
    \vspace{-40pt}
    \hbox to \hsize{ \footmsgA \hfill }
}
\makeatother

\def\evenmark{\bf entropy}
\def\oddmarkA{{\sl Entropy} {\bf 2004}{\sl , 6}}
\def\oddmarkB{, \thepagerange}
\def\oddmarkC{}
\def\oddmarkD{{\Huge \bf Entropy}}
\def\oddmarkE{\normalsize {\bf ISSN 1099-4300}}
\def\oddmarkF{{www.mdpi.org/entropy/}}
\def\footmsgA{{
}}

\newcommand{\lora} {\boldmath$\longrightarrow$}
\newcommand{\vecbm}[1]{\mbox{\boldmath#1}}
\newcommand{\vecb}[1]{\mbox{\bf#1}}
\newcommand{\cent}[1] {\begin{center}#1\end{center}}
\newcommand{\lra}  {$\leftrightarrow$}

\newcommand{\doublint}{\int\rule{-3.5mm}{0mm}\int}

\def\thedoctitle{\bf  A New Thermodynamics from Nuclei to Stars}

\def\theauthorname{Dieter.H.E. Gross}

\def\authoraddress{Hahn-Meitner Institute and Freie Universit{\"a}t Berlin,\\
Fachbereich Physik.\\ Glienickerstr. 100\\ 14109 Berlin, Germany
\\E-mail:
gross@hmi.de}

\def\thereceivedhistory{Received: 30 June 2003 / Accepted: 26 December 2003 / Published: 16 March 2004}

\begin{document}

{
    \noindent {\\ \\}
}

{
    \LARGE \noindent \thedoctitle
} \\

{
    \noindent {\bf {\theauthorname}}
} \\

{
    \normalsize \noindent {\authoraddress}

{ \noindent {\sl \thereceivedhistory} } \\

\vspace{2pt} \hbox to \hsize{\hrulefill}
\vspace{.1in}

\begin{inmargins}{1cm}
\noindent {\bf Abstract:} Equilibrium statistics of Hamiltonian systems is
correctly described
  by the microcanonical ensemble. Classically this is the manifold of all
  points in the $N-$body phase space with the given total energy.
  Due to Boltzmann's principle, $e^S=tr(\delta(E-H))$, its
  geometrical size is related to the entropy $S(E,N,\cdots)$. This
  definition does not invoke any information theory, no
  thermodynamic limit, no extensivity, and no homogeneity
  assumption, as are needed in conventional (canonical) thermo-statistics.
  Therefore, it describes the equilibrium statistics of extensive
  as well of non-extensive systems. Due to this fact it is the {\em
  fundamental} definition of any classical equilibrium statistics. It can
  address nuclei and astrophysical objects as well. All kind of
  phase transitions can be distinguished sharply and uniquely for even small
  systems. It is further shown that the second law is a natural consequence of the
  statistical nature of thermodynamics which describes all systems
  with the same -- redundant -- set of few control parameters simultaneously. It
  has nothing to do with the thermodynamic limit. It even works in
  systems which are by far {\em larger} than any thermodynamic "limit".\\

\noindent {\bf Keywords:} Classical Thermo-statistics, Non-extensive
systems.
\end{inmargins}

\vspace{2pt} \hbox to \hsize{\hrulefill}

\newpage


\def\oddmarkC{\thepage}
\def\oddmarkB{}
\def\oddmarkD{}
\def\oddmarkE{}
\def\oddmarkF{}
\def\footmsgA{}


\section{Introduction}
Classical thermodynamics and the theory of phase transitions of homogeneous
and large systems are some of the oldest and best established theories in
physics. It may look strange to add anything new to it. Let me recapitulate
what was told to us since $>100$ years and  still told today
c.f.\cite{goldstein03}:
\begin{itemize}
\item thermodynamics addresses large homogeneous systems at equilibrium
(in the thermodynamic limit $N\to\infty|_{ N/V=\rho, homogeneous}$).
\item Phase transitions are the positive zeros of the grand-canonical
partition sum $Z(T,\mu,V)$ as function of $e^{\beta\mu}$
(Yang-Lee-singularities). As the partition sum for a finite number of
particles is always positive, zeros can only exist in the thermodynamic
limit $V|_{\beta,\mu}\to\infty$.
\item Micro and canonical ensembles are equivalent.
\footnote{\label{foot} How does one normally prove this?: The general link
between the microcanonical probability $e^{S(E,N)}$ and the grand-canonical
partition function $Z(T,\mu)$ for extensive systems is by the Laplace
transform:
\begin{eqnarray} Z(T,\mu,V)&=&\doublint_0^{\infty}{\frac{dE}{\epsilon_0}\;
dN\;e^{-[E-\mu N-TS(E,N)]/T}}\\
&=&\frac{V^2}{\epsilon_0}\doublint_0^{\infty} {de\;dn\;e^{-V[e-\mu
n-Ts(e,n)]/T}}\label{grandsum}\\ &\approx&\hspace{2cm}e^{\mbox{
const.+lin.+quadr.}}\label{saddlepoint}
\end{eqnarray}\\~\\
If $s(e,n)$ is concave then there is a single point $e_s$,$n_s$ with
\begin{eqnarray}
\frac{1}{T}&=&\left.\frac{\partial S}{\partial E}\right|_s\nonumber\\
\frac{\mu}{T}&=&-\left.\frac{\partial S}{\partial N}\right|_s\nonumber
\end{eqnarray}
In the thermodynamic limit ($V\to\infty$) the quadratic approximation in
equ.(\ref{saddlepoint}) becomes exact and there is a one to one mapping of
the microscopic mechanical $e=E/V,n=N/V$ to the intensive $T,\mu$.}
\item thermodynamics works with intensive variables $T,P,\mu$.
\item Unique Legendre mapping $T\to E$.
\item Heat only flows from hot to cold (Clausius)
\item Second law only in infinite systems when the Poincarr\'{e}
recurrence time becomes infinite (much larger than the age of the universe
(Boltzmann)).
\end{itemize}

 Under these constraint only a tiny part of the real world of
equilibrium systems can be treated. The ubiquitous non-homogeneous systems:
nuclei, clusters, polymers, soft matter (biological) systems, but also the
largest, astrophysical systems are not covered. Even normal systems with
short-range coupling at phase separations are non-homogeneous and can only
be treated within conventional homogeneous thermodynamics (e.g.
van-der-Waals theory) by bridging the unstable region of negative
compressibility by an artificial Maxwell construction. Thus even the
original goal, for which thermodynamics was invented some $150$ years ago,
the description of steam engines is only artificially solved. There is no
(grand-)canonical ensemble of phase separated and, consequently,
non-homogeneous, configurations. It has a deep reason as I will discuss
below: here the systems have a {\em negative} heat capacity $C$ (resp.
susceptibility). This, however, is impossible in the (grand-)canonical
theory where $C\propto (\delta E)^2$

In this paper I will describe a generalization which takes Boltzmann's
principle serious and avoids the thermodynamic limit. This opens
thermodynamics to the much larger world of non-extensive systems. The most
prominent example are of course self-gravitating astrophysical systems
which I will discuss here also. All the so familiar paradigms above have an
only limited validity and are clearly violated at the most interesting
situations as shown below.

Moreover, it helps to clarify the {\em physical} background of the
statistical approach: The incompleteness of our knowledge of the degrees of
freedom leads to the various ensembles and to their characteristic
differences. {\em Thermodynamics describes the properties of the entire
ensemble not of a specific system (point in phase-space)}. Further, it will
clarify the importance and origin of the fluctuations. It will shed new
light on the microscopic mechanism leading to first order phase
transitions. Finally, it will illuminate the origin of irreversibility and
the second law out of the time-reversible microscopic equations of motion.
\section{Boltzmann's principle}
   The Microcanonical ensemble is the ensemble (manifold)
   of all possible points in the $6N$ dimensional phase space at
   the prescribed sharp energy $E$:
\begin{eqnarray}
W(E,N,V)&=&\epsilon_0 tr\delta(E-H_N)\\
tr\delta(E-H_N)&=&\int{\frac{d^{3N}p\;d^{3N}q}{N!(2\pi\hbar)^{3N}}
\delta(E-H_N)}\label{phasespintegr}.
\end{eqnarray}
thermodynamics addresses the whole ensemble. It is ruled by
  the topology of the geometrical size $W(E,N,\cdots)$,
 Boltzmann's principle:
 \begin{equation}
\fbox{\fbox{\vecbm{S=k*lnW}}}\label{boltzmentr1}
\end{equation}
is the most fundamental definition of the entropy $S$.  Entropy and with it
micro-canonical thermodynamics has therefore a pure mechanical, geometrical
foundation. No information theoretical formulation is needed. Moreover, in
contrast to the canonical entropy, $S(E,N,..)$ is everywhere single valued
and multiple differentiable. There is no need for extensivity, no need of
concavity, no need of additivity, and no need of the thermodynamic limit.
This is a great advantage of the geometric foundation of equilibrium
statistics over the conventional definition by the Boltzmann-Gibbs
canonical theory. However, addressing entropy to finite eventually small
systems we will face a new problem with Zermelo's objection against the
monotonic rise of entropy, the second law, when the system approaches its
equilibrium. This is discussed in section (\ref{secondL})
c.f.\cite{gross183,gross192}. A further comment: In contrast to many
authors like Schr\"odinger \cite{schroedinger46} our ensemble is {\em not}
an ensemble of non-interacting replica of the considered system which may
exchange energy. I do {\em not} consider the different ways to distribute
energy over the different replica. I consider the manifold of the same
system at the precisely given energy under all possible different
distributions of the momenta and positions of its constituents (particles)
in the $6N$-dimensional phase space. The result is then the average
behaviour when one does not know the precise position and momentum of every
particle but only the total energy.
\section{Topological properties of $S(E,\cdots)$}

The topology of the Hessian of $S(E,\cdots)$, the determinant of curvature
of $s(e,n)$ determines completely all kinds phase transitions. This is
evidently so, because $e^{S(E)-E/T}$ is the weight of each energy in the
canonical partition sum at given $T$, see footnote \ref{foot}.
Consequently, at phase separation this has at least two maxima, the two
phases. And in between two maxima there must be a minimum where the
curvature of $S(E)$ is positive. I.e. the positive curvature detects phase
separation. This is also in the case of several conserved control
parameters.
\begin{eqnarray}
d(e,n)&=&\left\|\begin{array}{cc} \frac{\partial^2 s}{\partial e^2}&
\frac{\partial^2 s}{\partial n\partial e}\\ \frac{\partial^2 s}{\partial
e\partial n}& \frac{\partial^2 s}{\partial n^2}
\end{array}\right\|=\lambda_1\lambda_2 \label{curvdet}\\
\lambda_1&\ge&\lambda_2\hspace{1cm}\mbox{\lora eigenvectors :}\hspace{1cm}
{\boldmath\vecbm{$v$}_1,\vecbm{$v$}_2}\nonumber
\end{eqnarray}
Of course for a finite system each of these maxima of $S(E,\cdots)-E/T$
have a non-vanishing width. There are intrinsic fluctuations in each phase.
\subsection{ Unambiguous signal of phase transitions in a "Small"
system \cite{gross170}} The whole zoo of phase-transitions can be sharply
seen and distinguished. This is here demonstrated for the Potts-gas model
on a two dimensional lattice of {\em finite} size of $50\times 50$ lattice
points, c.f. fig.(\ref{det}).
\begin{figure}[h]
\includegraphics*[bb =0 0 290 180, angle=0, width=12cm,
clip=true]{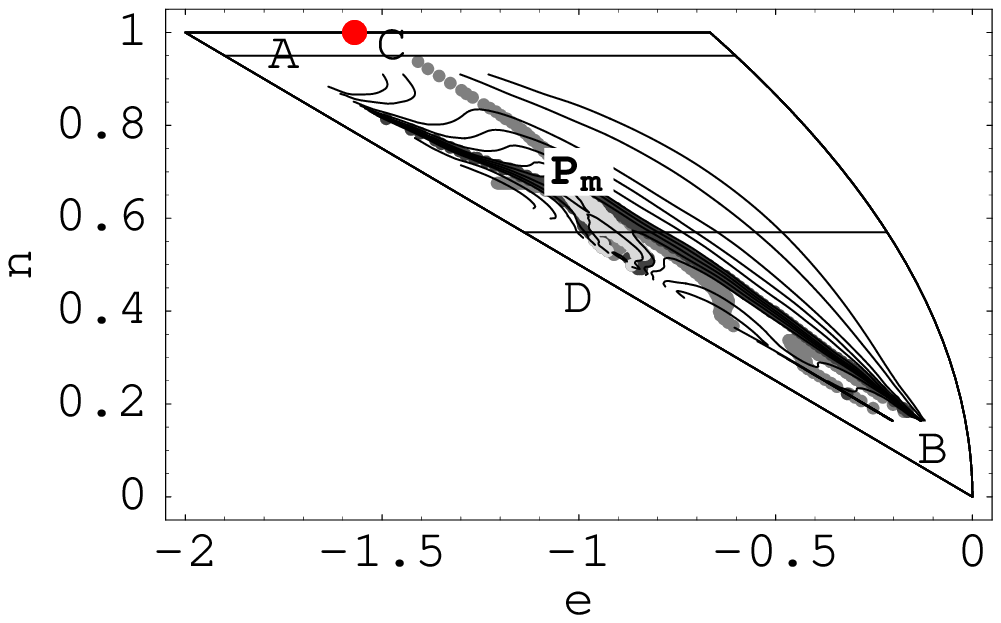}
\caption{Global phase diagram or contour plot of the curvature determinant
  (Hessian), eqn.~(\ref{curvdet}), of the 2-dim Potts-3 lattice gas
  with $50*50$ lattice points, $n$ is the number of particles per
  lattice point, $e$ is the total energy per lattice point. The line
   (-2,1) to (0,0) is the ground-state energy of the lattice-gas
   as function of $n$. The most right curve is the locus of configurations
    with completely random spin-orientations (maximum entropy). The whole
    physics of the model plays between these two boundaries.  At the
    dark-gray lines the Hessian is $\det=0$,this is the boundary of the
    region of phase separation (the triangle $AP_mB$) with a negative
    Hessian ($\lambda_1>0,\lambda_2<0$).  Here, we have
    Pseudo-Riemannian geometry. At the light-gray lines
    is a minimum of $\det(e,n)$ in the direction of the largest
    curvature (\vecbm{v}$_{\lambda_{max}}\cdot$\vecbm{$\nabla$}$\det=0$) and
    $\det=0$,these are lines of second order transition. In the triangle
    $AP_mC$ is the pure ordered (solid) phase ($\det>0, \lambda_1<0$).
    Above and right of the line $CP_mB$ is the pure disordered (gas) phase
    ($\det>0, \lambda_1<0$). The crossing $P_m$ of the boundary lines is a
    multi-critical point. It is also the critical end-point of the region
    of phase separation ($\det<0,\lambda_1>0,\lambda_2<0$).  The light-gray
    region around the multi-critical point $P_m$ corresponds to a flat,
    horizontal region of $\det(e,n)\sim 0$ and consequently to a somewhat
    extended cylindrical region of $s(e,n)$ and {\boldmath
    \vecbm{$\nabla$}\mbox{$\lambda_1$}{\boldmath$\sim 0$}},
    details see \protect\cite{gross173,gross174}; $C$ is the analytically
   known position of the critical point which the ordinary $q=3$ Potts
   model (without vacancies){\em would have in the thermodynamic
   limit}}\label{det}
\end{figure}
\subsection{Systematic of phase transitions in the micro-canonical ensemble without
invoking the thermodynamic limit}

\begin{itemize}
\item A single stable phase of course with some intrinsic fluctuations (width) by
a negative largest curvature $\lambda_1<0$. Here $s(e,n)$ is concave
(downwards bending) in both directions.  Then there is a one to one mapping
of the canonical \lra the micro-ensemble. \cent{\includegraphics*[bb = 53
14 430 622, angle=-90, width=9cm, clip=true]{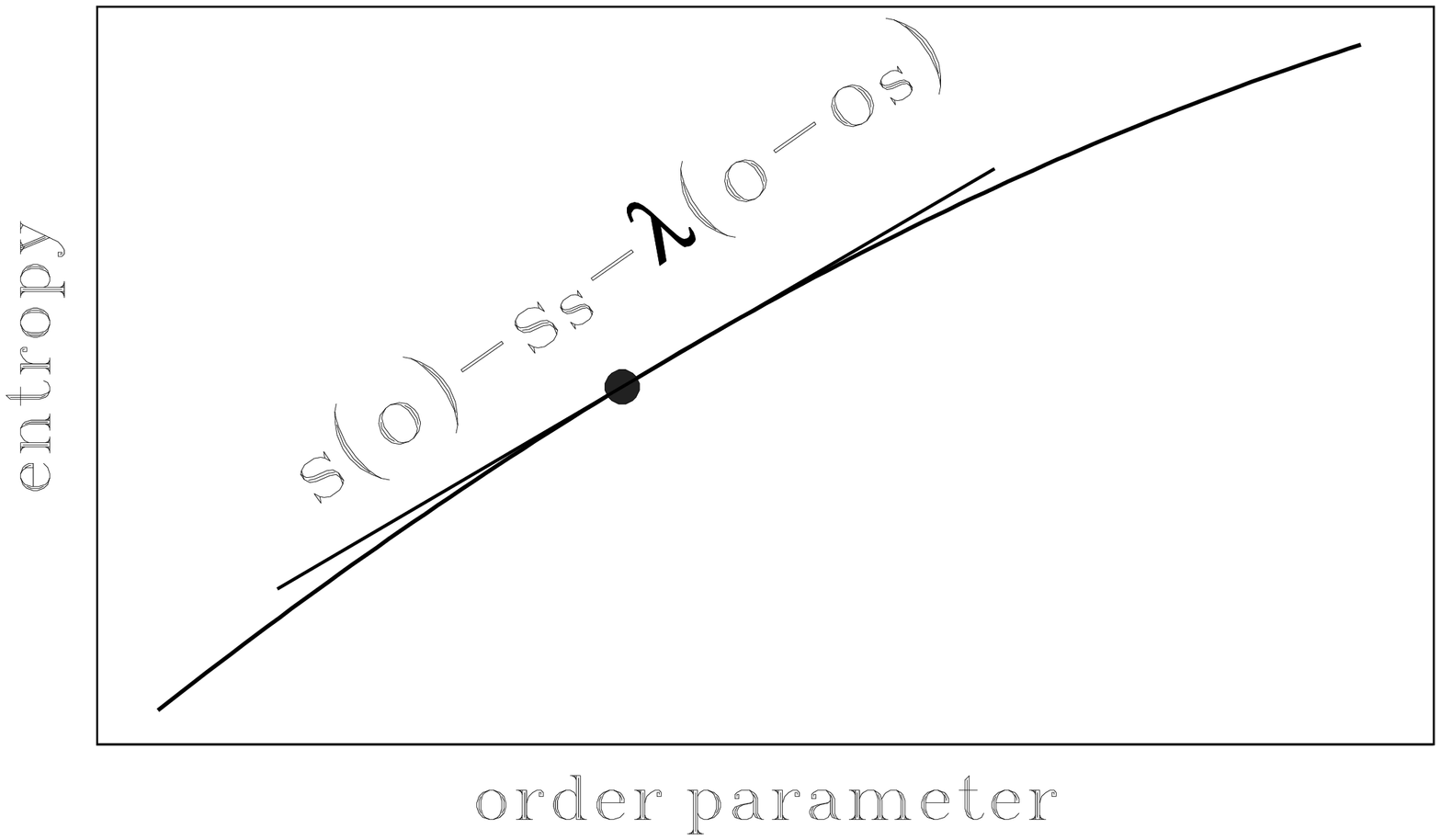}}
\item A transition of first order with phase separation
  and surface tension is indicated by
$\lambda_1(e,n)>0$. $s(e,n)$ has a convex intruder (upwards bending) in the
direction
  $\vecb{v}_1$ of the largest curvature $\ge 0$  which can be identified with the
  order parameter \cite{gross174}. Three solutions of
\begin{eqnarray}
\beta=\frac{1}{T}&=&\left.\frac{\partial S}{\partial E}\right|_s\nonumber\\
\nu=\frac{\mu}{T}&=&-\left.\frac{\partial S}{\partial N}\right|_s\nonumber
\end{eqnarray} determine the intensive temperature $T=1/\beta$ and
the chemical potential $T\nu$.
\cent{\includegraphics*[bb = 105 15 463 623, angle=-90, width=9cm,
  clip=true]{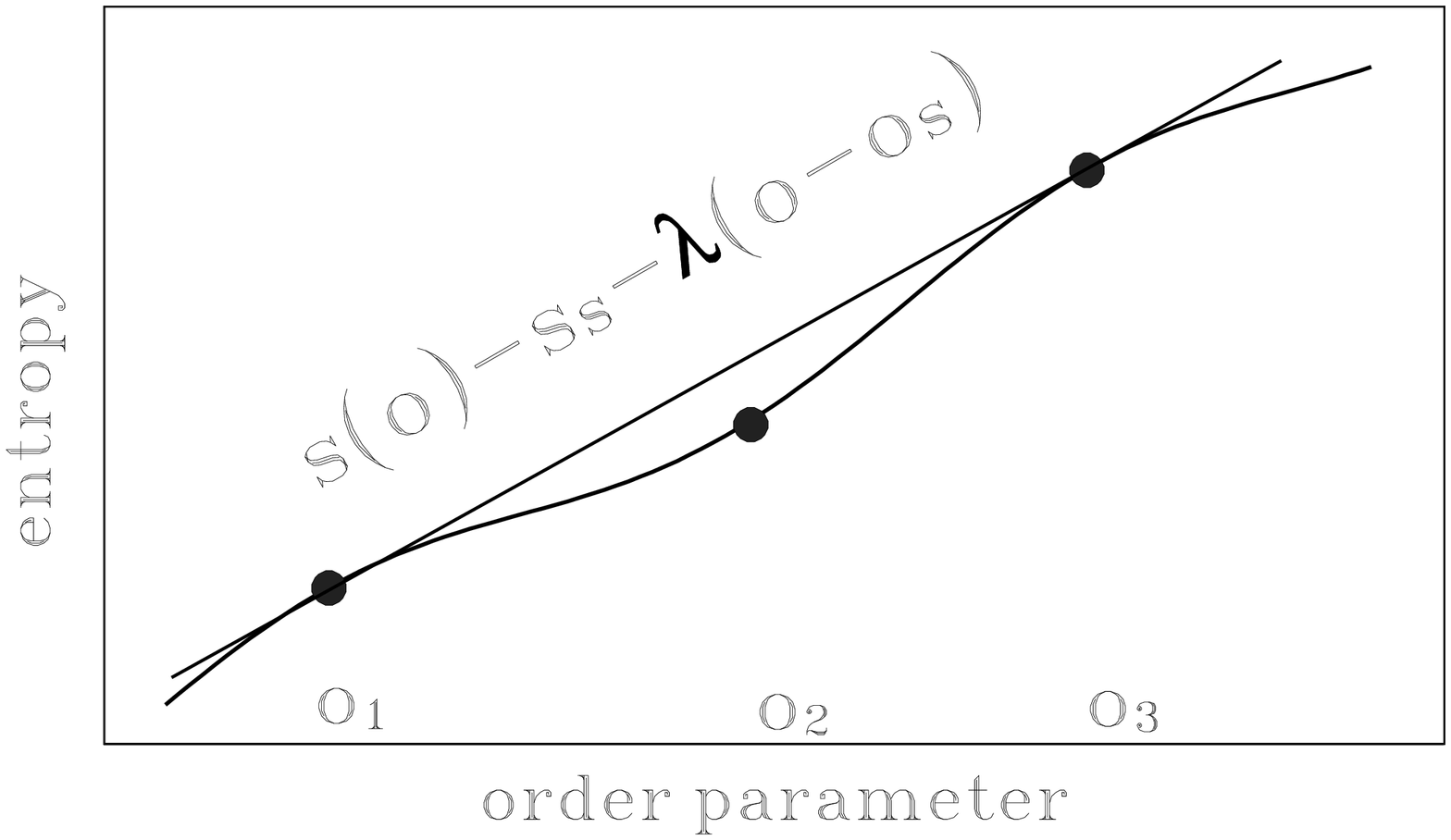}}
In the thermodynamic limit the whole region \{$o_1,o_3$\} is mapped into a
single point in the canonical ensemble which is consequently non-local in
$o$. I.e. if the curvature of $S(E,N)$ is $\lambda_1\ge 0$ both ensembles
are not equivalent even in the limit.
\item A continuous (``second order'') transition
with vanishing surface tension, where two neighboring phases become
indistinguishable. This is indicated in figure (\ref{det}) by a line with
$\lambda_1=0$ and extremum of $\lambda_1$ in the direction of order
parameter $ \vecb{v}_{\lambda=0}\cdot\vecb{$\nabla$}\lambda_1=0$. These are
the catastrophes of the Laplace transform $E\to T$.
\end{itemize}
\subsection{CURVATURE}

We saw that the curvature (Hessian) of $S(E,N,\cdots)$ controls the phase
transitions. What is the physics behind the curvature? For short-range
force it is linked to the interphase surface tension.

\begin{table}[h]
\caption{Parameters of the liquid--gas transition of small
  sodium clusters (MMMC-calculation~\protect\cite{gross174}) in
  comparison with the bulk for a rising number $N_0$ of atoms,
  $N_{surf}$ is the average number of surface atoms (estimated here as
  $\sum{N_{cluster}^{2/3}}$) of all clusters with $N_i\geq2$ together.
  $\sigma/T_{tr}=\Delta s_{surf}*N_0/N_{surf}$ corresponds to the
  surface tension. Its bulk value is adjusted
  to agree with the experimental values of the $a_s$ parameter which
  we used in the liquid-drop formula for the binding energies of small
  clusters, c.f.  Brechignac et al.~\protect\cite{brechignac95}, and
  which are used in this calculation~\cite{gross174} for the individual
  clusters.}
\begin{center}
\renewcommand{\arraystretch}{1.4}
\setlength\tabcolsep{5pt}
\begin{tabular} {|c|c|c|c|c|c|} \hline
&$N_0$&$200$&$1000$&$3000$&\vecb{bulk}\\ 
\hline \hline &$T_{tr} \;[K]$&$940$&$990$&$1095$&\vecb{1156}\\ \cline{2-6}
&$q_{lat} \;[eV]$&$0.82$&$0.91$&$0.94$&\vecb{0.923}\\ \cline{2-6} {\bf
Na}&$s_{boil}$&$10.1$&$10.7$&$9.9$&\vecb{9.267}\\ \cline{2-6} &$\Delta
s_{surf}$&$0.55$&$0.56$&$0.44$&\\ \cline{2-6}
&$N_{surf}$&$39.94$&$98.53$&$186.6$&\vecbm{$\infty$}\\ \cline{2-6}
&$\sigma/T_{tr}$&$2.75$&$5.68$&$7.07$&\vecb{7.41}\\ \hline
\end{tabular}
\end{center}
\end{table}
\begin{figure}[h]\cent{
\includegraphics*[bb = 99 57 400 286, angle=-0, width=9cm,
clip=true]{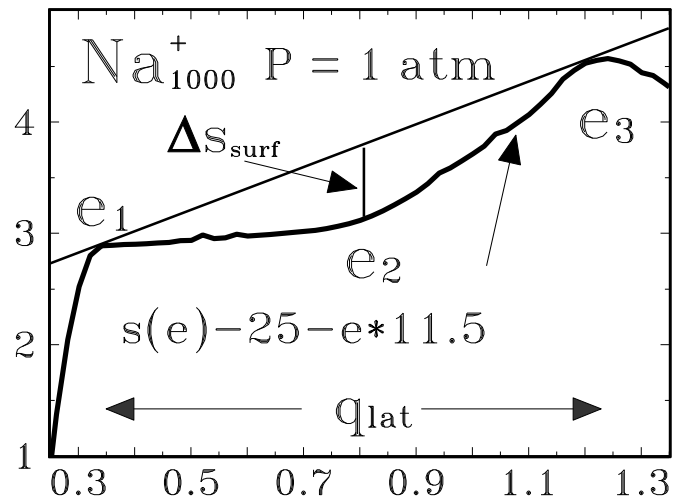}} \caption{MMMC~\protect\cite{gross174} simulation
of the entropy $s(e)$ per atom ($e$ in eV per atom) of a system of
  $N_0=1000$ sodium atoms at an external pressure of 1 atm.  At the
  energy $e\leq e_1$ the system is in the pure liquid phase and at
  $e\geq e_3$ in the pure gas phase, of course with fluctuations. The
  latent heat per atom is $q_{lat}=e_3-e_1$.  \underline{Attention:}
  the curve $s(e)$ is artificially sheared by subtracting a linear
  function $25+e*11.5$ in order to make the convex intruder visible.
  {\em $s(e)$ is always a steep monotonic rising function}.  We
  clearly see the global concave (downwards bending) nature of $s(e)$
  and its convex intruder. Its depth is the entropy
  loss due to additional correlations by the interfaces. It scales $\propto
  N^{-1/3}$. From this one can calculate the surface
  tension per surface atom
  $\sigma_{surf}/T_{tr}=\Delta s_{surf}*N_0/N_{surf}$.  The double
  tangent (Gibbs construction) is the concave hull of $s(e)$. Its
  derivative gives the Maxwell line in the caloric curve $T(e)$ at
  $T_{tr}$. In the thermodynamic limit the intruder would disappear and $s(e)$
  would approach the double tangent from below.  Nevertheless, even
  there, the probability $\propto e^{Ns}$ of configurations with
  phase-separations are suppressed by the
  (infinitesimal small) factor $e^{-N^{2/3}}$ relative to the pure
  phases and the distribution remains {\em strictly bimodal in the
    canonical ensemble}. The region $e_1<e<e_3$ of phase separation
  gets lost.\label{naprl0f}}
\end{figure}
\clearpage
\subsection{Heat can flow from cold to hot}
In figure (\ref{heat}) it is shown how a convexity of $s(e)$ leads to a
violation of Clausius' first formulation of the second law.
\begin{figure}[h]
\begin{center}
\includegraphics*[bb =38 8 387 611, angle=-180, width=6.7 cm,
clip=true]{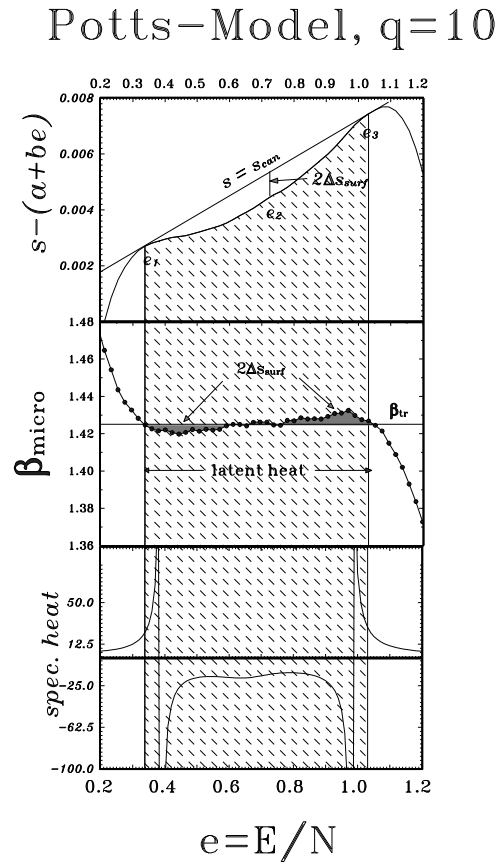}\caption{ Potts model, ($q=10$) in the region of phase
separation. At $e_1$ the system is in the pure ordered phase, at $e_3$ in
the pure disordered phase. A little above $e_1$ the temperature $T=1/\beta$
is higher than a little below $e_3$. Combining two parts of the system: one
at the energy $e_1+\delta e$ and at the temperature $T_1$, the other at the
energy $e_3-\delta e$ and at the temperature $T_3<T_1$ will equilibrize
with a rise of its entropy, a drop of $T_1$ (cooling) and an energy flow
(heat) from $3\to 1$: i.e.: Heat flows from cold to hot! Clausius
formulation of the second law is violated. Evidently, this is not any
peculiarity of gravitating systems! This is a generic situation within
classical thermodynamics even of systems with short-range coupling and {\em
has nothing to do with long range interaction.}\label{heat}}\vspace{-2cm}
\end{center}
\end{figure}
\clearpage
\section{ Negative heat capacity as signal for a phase
transition of first order.} In the previous discussion we saw the
phenomenon of first-order phase-transitions is linked to the appearance of
non-homogeneities at phase-separations, and inter-phase surfaces. This,
then gives rise to the convexity of $S(\mbox{order-parameter}$ and the
appearance of negative susceptibilities (e.g. a negative heat capacity).
{\em These are ubiquitous in nature from the smallest to the largest
systems:}
\subsection{Nuclear Physics}
A very detailed illustration of the appearance of negative heat capacities
is given by D'Agostino et al. \cite{dAgostino00}. Here I want to remember
one of the oldest experimental finding of a "back"-bending caloric curve in
Nuclear Physics.
\begin{figure}[h]
\includegraphics*[bb = 8 29 495 597, angle=-90, width=10 cm,  
clip=true ]{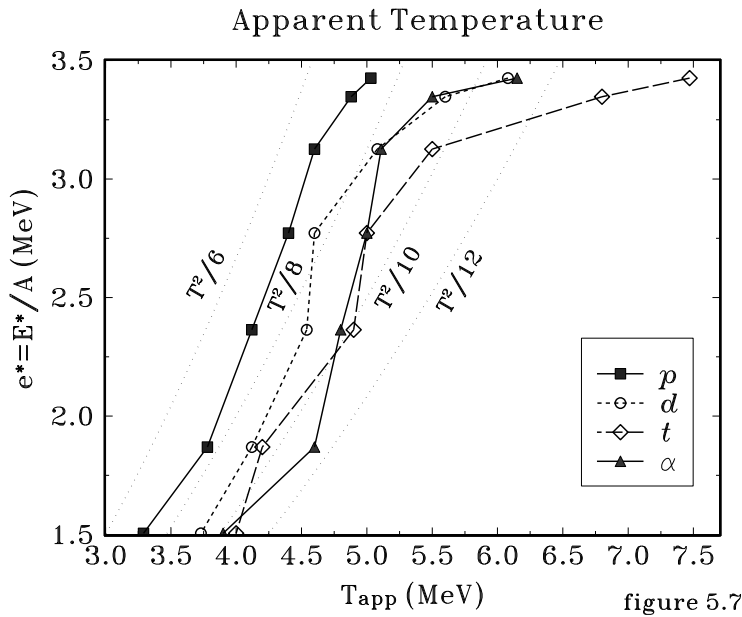}                                         
\caption{Experimental excitation energy per nucleon $e^*$ versus
  apparent temperature $T_{app}$ for backward $p$,
  $d$, $t$ and $\alpha$ together with heavy evaporation residues out of
  incomplete fusion of  $701$ Mev $^{28}$Si$+^{100}$Mo. The dotted curves give the
  Fermi-gas caloric curves for the level-density
  parameter a = 6 to 12. (Chbihi et al. Eur.Phys.J. A 1999)}
\end{figure}
\clearpage
\subsection{Atomic clusters}
Here I show the simulation of a typical fragmentation transition of a
system of $3000$ sodium atoms interacting by realistic (many-body) forces.
To compare with usual macroscopic conditions, the calculations were done at
each energy using a volume $V(E)$ such that the microcanonical pressure
$P=\frac{\partial S}{\partial V}/\frac{\partial S}{\partial E}=1$atm. The
inserts above give the mass distribution at the various points. The label
"4:1.295" means 1.295 quadrimers on average.  This gives a detailed insight
into what happens with rising excitation energy over the transition region:
At the beginning ($e^*\sim 0.442$ eV) the liquid sodium drop evaporates 329
single atoms and 7.876 dimers and 1.295 quadrimers on average. At energies
$e>\sim 1$eV the drop starts to fragment into several small droplets
("intermediate mass fragments") e.g. at point 3: 2726 monomers,80
dimers,$\sim$10 trimers, $\sim$30 quadrimers and a few heavier ones up to
10-mers. The evaporation residue disappears. This multifragmentation
finishes at point 4. It induces the strong backward swing of the caloric
curve $T(E)$. Above point 4 one has a gas of free monomers and at the
beginning a few dimers. This transition scenario has a lot similarity with
nuclear multifragmentation. It is also shown how the total interphase
surface area, proportional to $N_{eff}^{2/3}=\sum_i N_i^{2/3}$ with $N_i\ge
2$ ($N_i$ the number of atoms in the $i$th cluster) stays roughly constant
up to point 3 even though the number of fragments ($N_{fr}=\sum_i$) is
monotonic rising with increasing excitation.
\begin{figure}[h]\vspace*{-0.5
cm}
 \cent{\includegraphics [bb = 69 382 545
766, angle=-0, width=12cm, clip=true]{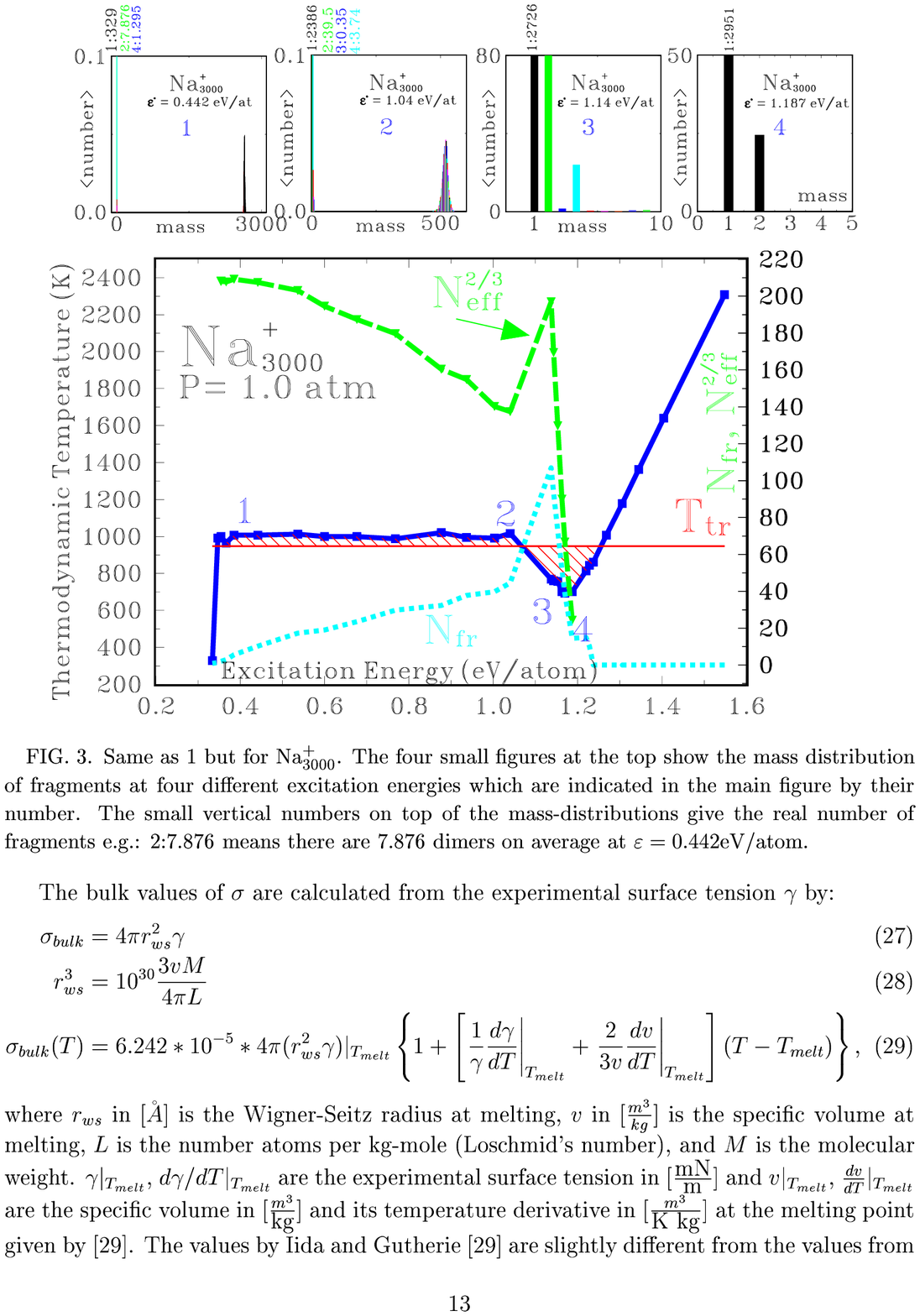}} \vspace*{-0.5
cm}\caption{Cluster fragmentation}
\end{figure}
\clearpage

\subsection{Stars}
    Self-gravitation leads to a non-extensive potential energy $\propto
N^2$.  No thermodynamic limit exists for $E/N$ and no canonical treatment
makes sense. At negative total energies these systems have a negative heat
capacity.  This was for a long time considered as an absurd situation
within canonical statistical mechanics with its thermodynamic ``limit''.
However, within our geometric theory this is just a simple example of the
pseudo-Riemannian topology of the microcanonical entropy $S(E,N)$ provided
that high densities with their non-gravitational physics, like nuclear
hydrogen burning, are excluded. We treated the various phases of a
self-gravitating cloud of particles as function of the total energy and
angular momentum, c.f. the quoted paper. Clearly these are the most
important constraint in astrophysics.
    \begin{figure}[h]
    \includegraphics*[bb =88 401 522 630, angle=-0, width=12 cm,
    clip=true]{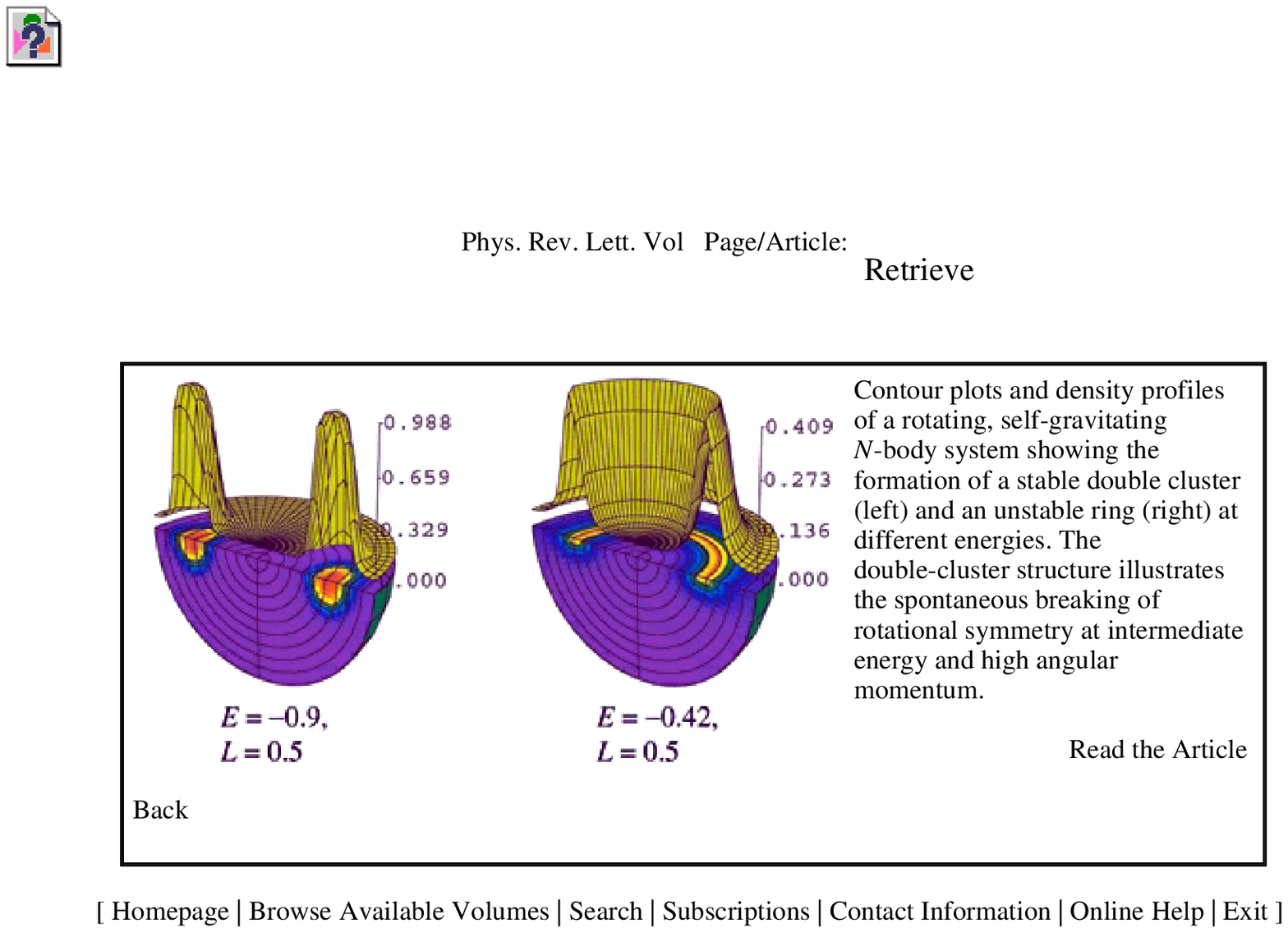}
    \caption{Phases and Phase-Separation
    in Rotating, Self-Gravitating Systems,
    Physical Review Letters--July 15, 2002, cover-page, by
    (Votyakov, Hidmi, De Martino, Gross)}
    \end{figure}
\begin{figure}
\includegraphics[bb =72 54 533 690,width=8cm,angle=-90,clip=true]{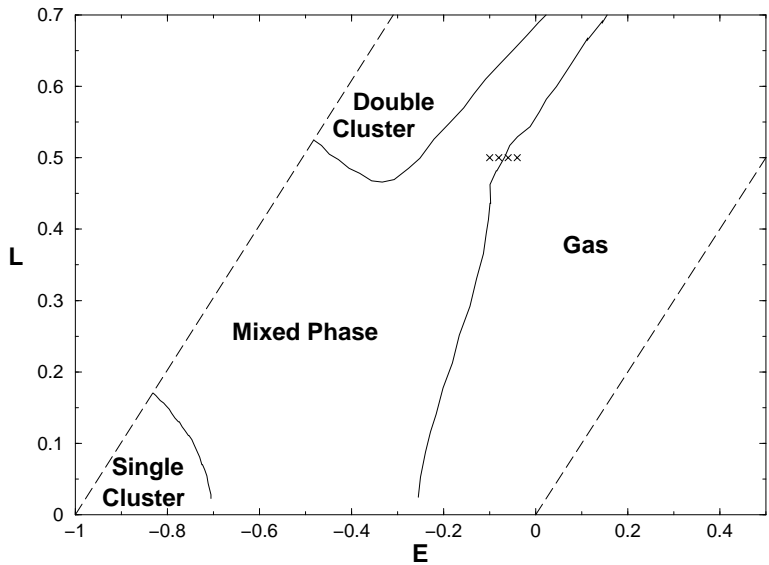}
\caption{Microcanonical phase-diagram of a cloud of self-gravitating and
rotating system as function of the energy and angular-momentum. Outside the
dashed boundaries only some singular points were calculated. In the mixed
phase the largest curvature $\lambda_1$ of $S(E,L)$ is positive.
Consequently the heat capacity or the correspondent susceptibility is
negative. This is of course well known in astrophysics. However, the new
and important point of our finding is that within microcanonical
thermodynamics this is {\em a generic property of all phase transitions of
first order, independently of whether there is a short- or a long-range
force that organizes the system}.}
\end{figure}

\section{The second law, microcanonically\label{secondL}}
There are many attempts to derive the second law of thermodynamics from
Statistical Mechanics e.g. by Einstein or recently by Lieb and many others
\cite{einstein02,einstein03,lieb97,lieb98a}. Lebowitz gave an excellent and
deep discussion of the present day understanding of irreversibility
\cite{lebowitz99,lebowitz99a,goldstein03}. His ``Statistical mechanics: A
selective review of two central issues'' addresses exactly to the two
issues I want to discuss here. They are closely related but there are also
deep differences as will become evident in the following.

In this paper I want to emphasize an important aspect of statistical
mechanics of many-body systems which to my opinion was not sufficiently
considered up to now: Statistical Mechanics and with it also thermodynamics
are macroscopic theories describing the {\em average} behaviour of {\em
all} many-body systems with the same but {\em redundant} macroscopic
constraint and information. This fact leads in a simple and straight
forward manner to the desired understanding of irreversibility and the
second law for systems obeying a time reversible microscopic dynamics. It
is important to deduce the second law from reversible (here Newtonian) and
not from dissipative dynamics as is often done because just the {\em
derivation of irreversibility} from fully {\em reversible} dynamics is the
main challenge.

The new approach that I will present here was derived from the experience
that Gibbs' canonical statistics fails to describe properly the large and
important group of non-extensive systems. These are systems subjected to an
interaction with a range comparable to or larger than their linear
dimension. They are strongly non-homogeneous and are either finite and have
a surface, or they show a strong clustering like astrophysical objects, or
they are split spatially in regions of different structures (phases) like a
system at the liquid-gas phase separation.

\subsection{Measuring a macroscopic observable, the ``
EPS-formulation ''} \label{EPSformulation} A single point
$\{q_i(t),p_i(t)\}_{i=1\cdots N}$ in the $N$-body phase space corresponds
to a detailed specification of the system with all degrees of freedom
(d.o.f) completely fixed at time $t$ (microscopic determination).  Fixing
only the total energy $E$ of an $N$-body system leaves the other
($6N-1$)-degrees of freedom unspecified.  A second system with the same
energy is most likely not in the same microscopic state as the first, it
will be at another point in phase space, the other d.o.f. will be
different. I.e. the measurement of the total energy $\hat{H}_N$, or any
other macroscopic observable $\hat{M}$, determines a ($6N-1$)-dimensional
{\em sub-manifold} ${\cal{E}}$ or ${\cal{M}}$ in phase space\footnote{In
this paper we
  denote ensembles or manifold in phase space by calligraphic letters
  like ${\cal{M}}$}. (The manifold ${\cal{M}}$ is called by Lebowitz a
{\em macrostate}\cite{lebowitz99,lebowitz99a,goldstein03} which contains
$\Gamma_M=W(M)$ microstates.) All points (the microstates) in $N$-body
phase space consistent with the given value of $E$ and volume $V$, i.e. all
points in the ($6N-1$)-dimensional sub-manifold ${\cal{E}}(N,V)$ of phase
space are equally consistent with this measurement. ${\cal{E}}(N,V)$ is the
microcanonical ensemble. This example tells us that {\em any macroscopic
measurement is incomplete
  and defines a sub-manifold of points in phase space not a single
  point}. An additional measurement of another macroscopic quantity
$\hat{B}\{q,p\}$ reduces ${\cal{E}}$ further to the cross-section
${\cal{E}\cap\cal{B}}$, a ($6N-2$)-dimensional subset of points in
${\cal{E}}$ with the volume:
\begin{equation}
W(B,E,N,V)=\frac{1}{N!}\int{\left(\frac{d^3q\;d^3p}
{(2\pi\hbar)^3}\right)^N\epsilon_0\delta(E-\hat H_N\{q,p\})\; \delta(B-\hat
B\{q,p\})} \label{integrM}\end{equation} If $\hat H_N\{q,p\}$ as also $\hat
B\{q,p\}$ are continuous differentiable functions of their arguments, what
we assume in the following, ${\cal{E}}\cap{\cal{B}}$ is closed. In the
following we use $W$ for the Riemann or Liouville volume of a many-fold.

Microcanonical thermo{\em statics} gives the probability $P(B,E,N,V)$ to
find the $N$-body system in the sub-manifold ${\cal{E}\cap\cal{B}}(E,N,V)$:
\begin{equation}
P(B,E,N,V)=\frac{W(B,E,N,V)}{W(E,N,V)}=e^{\ln[W(B,E,N,V)]-S(E,N,V)} \label{
EPS }\end{equation} This is what Krylov seems to have had in mind
\cite{krylov79} and what I will call the ``ensemble probabilistic
formulation of statistical mechanics ($EPS$) ''.

Similarly thermo{\em dynamics} describes the development of some
macroscopic observable $\hat{B}\{q_t,p_t\}$ in time of systems which were
specified at an earlier time $t_0$ by another macroscopic measurement
$\hat{A}\{q_0,p_0\}$.  It is related to the volume of the sub-manifold
${\cal{M}}(t)={\cal{A}}(t_0)\cap{\cal{B}}(t)\cap{\cal{E}}$:
\begin{equation} W(A,B,E,t)=\frac{1}{N!}\int{\left(\frac{d^3q_t\;d^3p_t}
{(2\pi\hbar)^3}\right)^N\delta(B-\hat B\{q_t,p_t\})\; \delta(A-\hat
A\{q_0,p_0\})\;\epsilon_0\delta(E-\hat H\{q_t,p_t\})}, \label{wab}
\end{equation}
where $\{q_t\{q_0,p_0\},p_t\{q_0,p_0\}\}$ is the set of trajectories
solving the Hamilton-Jacobi equations
\begin{equation}
\dot{q}_i=\frac{\partial\hat H}{\partial p_i},\hspace{1cm}
\dot{p}_i=-\frac{\partial\hat H}{\partial q_i},\hspace{1cm}i=1\cdots N
\end{equation}
with the initial conditions $\{q(t=t_0)=q_0;\;p(t=t_0)=p_0\}$. For a very
large system with $N\sim 10^{23}$ the probability to find a given value
$B(t)$, $P(B(t))$, is usually sharply peaked as function of $B$. Ordinary
thermodynamics treats systems in the thermodynamic limit $N\to\infty$ and
gives only $<\!\!B(t)\!\!>$. However, here we are interested to formulate
the second law for ``Small'' systems i.e.  we are interested in the whole
distribution $P(B(t))$ not only in its mean value $<\!\!B(t)\!\!>$.
thermodynamics does {\em not} describe the temporal development of a {\em
single} system (single point in the $6N$-dim phase space).

There is an important property of macroscopic measurements: Whereas the
macroscopic constraint $\hat{A}\{q_0,p_0\}$ determines (usually) a compact
region ${\cal{A}}(t_0)$ in \{$q_0,p_0$\} this does not need to be the case
at later times $t\gg t_0$: ${\cal{A}}(t)$ defined by
${\cal{A}}\{q_0\{q_t,p_t\},p_0\{q_t,p_t\}\}$ might become a {\em
  fractal} i.e. ``spaghetti-like'' manifold as a function of
$\{q_t,p_t\}$ in ${\cal{E}}$ at $t\to \infty$ and loose compactness.

This can be expressed in mathematical terms: There exist series of points
$\{a_n\}\in{\cal{A}}(t)$ which converge to a point $a_\infty$ which is {\em
not} in ${\cal{A}}(t)$. E.g. such points $a_\infty$ may have intruded from
the phase space complimentary to ${\cal{A}}(t_0)$. Illustrative examples
for this evolution of an initially compact sub-manifold into a fractal set
are the baker transformation discussed in this context by ref.
\cite{fox98,gilbert00}. Then no macroscopic (incomplete) measurement at
late time $t$ can resolve $a_\infty$ from its immediate neighbors $a_n$ in
phase space with distances $|a_n-a_\infty|$ less then any arbitrary small
$\delta$. In other words, {\em at the time $t\gg t_0$ no macroscopic
measurement with its  incomplete information about $\{q_t,p_t\}$ can decide
whether  $\{q_0\{q_t,p_t\},p_0\{q_t,p_t\}\}\in{\cal{A}}(t_0)$ or not.} I.e.
any macroscopic theory like thermodynamics can only deal with the {\em
closure} of ${\cal{A}}(t)$. If necessary, the sub-manifold ${\cal{A}}(t)$
must be artificially closed to $\overline{\cal{A}}(t)$ as developed further
in section \ref{fractSL}.  {\em Clearly, in this
  approach this is the physical origin of irreversibility.}
\subsection{Fractal distributions in phase space, second
law}\label{fractSL} Here we will first describe a simple working-scheme
(i.e. a sufficient method) which allows to deduce mathematically the second
law. Later, we will show how this method is necessarily implied by the
reduced information obtainable by macroscopic measurements.

Let us examine the following Gedanken experiment: Suppose the probability
to find our system at points $\{q_t,p_t\}_1^N$ in phase space is uniformly
distributed for times $t<t_0$ over the sub-manifold ${\cal{E}}(N,V_1)$ of
the $N$-body phase space at energy $E$ and spatial volume $V_1$. At time
$t>t_0$ we allow the system to spread over the larger volume $V_2>V_1$
without changing its energy.  If the system is {\em dynamically mixing},
the majority of trajectories $\{q_t,p_t\}_1^N$ in phase space starting from
points $\{q_0,p_0\}$ with $q_0\subset V_1$ at $t_0$ will now spread over
the larger volume $V_2$.  Of course the Liouvillean measure of the
distribution ${\cal{M}}\{q_t,p_t\}$ in phase space at $t>t_0$ will remain
the same ($=tr[{\cal{E}}(N,V_1)]$) \cite{goldstein59}. (The label
$\{q_0\subset V_1\}$ of the integral means that the positions $\{q_0\}_1^N$
are restricted to the volume $V_1$, the momenta $\{p_0\}_1^N$ are
unrestricted.)
\begin{eqnarray}
\left.tr[{\cal{M}}\{q_t\{q_0,p_0\},p_t\{q_0,p_0\}\}] \right|_{\{q_0\subset
V_1\}} &=&\int_{\{q_0\{q_t,p_t\}\subset
V_1\}}{\frac{1}{N!}\left(\frac{d^3q_t\;d^3p_t}
{(2\pi\hbar)^3}\right)^N\epsilon_0\delta(E-\hat H_N\{q_t,p_t\})}\nonumber\\
&=&\int_{\{q_0\subset V_1\}}{\frac{1}{N!}\left(\frac{d^3q_0\;
d^3p_0}{(2\pi\hbar)^3}\right)^N\epsilon_0\delta(E-\hat H_N\{q_0,p_0\})},\\
\mbox{because of: }\frac{\partial\{q_t,p_t\}}{\partial\{q_0,p_0\}}&=&1.
\end{eqnarray}
But as already argued by Gibbs the distribution ${\cal{M}}\{q_t,p_t\}$ will
be filamented like ink in water and will approach any point of
${\cal{E}}(N,V_2)$ arbitrarily close. ${\cal{M}}\{q_t,p_t\}$ becomes dense
in the new, larger ${\cal{E}}(N,V_2)$ for times sufficiently larger than
$t_0$.  The closure $\overline{{\cal{M}}}$ becomes equal to
${\cal{E}}(N,V_2)$.

In order to express this fact mathematically, {\em we have to redefine
  Boltzmann's definition of entropy eq.(\ref{boltzmentr1}) and
  introduce the following fractal ``measure'' for integrals like
  (\ref{phasespintegr}) or (\ref{integrM}):}
\begin{equation}
W(E,N,t\gg t_0)= \frac{1}{N!}\int_{\{q_0\{q_t,p_t\}\subset
V_1\}}{\left(\frac{d^3q_t\;d^3p_t}
{(2\pi\hbar)^3}\right)^N\epsilon_0\delta(E-\hat H_N\{q_t,p_t\})}
\end{equation}
With the transformation (done in appropriate dimension-less units) :
\begin{eqnarray}
\int{\left(d^3q_t\;d^3p_t\right)^N\cdots}&=& \int{d\sigma_1\cdots
d\sigma_{6N}\cdots}\\ d\sigma_{6N}&:=&\frac{1}{||\nabla\hat H||}
\sum_i{\left(\frac{\partial\hat H}{\partial q_i}dq_i+\frac{\partial\hat
H}{\partial p_i}dp_i\right)}=
 \frac{1}{||\nabla\hat H||}dE\\
||\nabla\hat H||&=&\sqrt{\sum_i{\left(\frac{\partial\hat H}{\partial
q_i}\right)^2+\sum_i{\left(\frac{\partial\hat H}{\partial
p_i}\right)^2}}}\\ W(E,N,t\gg t_0)&=&\frac{1}{N!(2\pi\hbar)^{3N}}
\int_{\{q_0\{q_t,p_t\}\subset V_1\}} {d\sigma_1\cdots d\sigma_{6N-1}
\frac{\epsilon_0}{||\nabla\hat H||}},
\end{eqnarray}
we replace ${\cal{M}}$ by its closure $\overline{\cal{M}}$ and {\em
  define} now:
\begin{equation}
W(E,N,t\gg t_0)\to M(E,N,t\gg t_0):=<\!G({\cal{E}}(N,V_2))\!>
*\mbox{vol}_{box}[{\cal{M}}(E,N,t\gg t_0)],\label{boxM1}
\end{equation}
where $<\!G({\cal{E}}(N,V_2))\!>$ is the average of
$\frac{\epsilon_0}{N!(2\pi\hbar)^{3N}||\nabla\hat H||}$ over the (larger)
manifold ${\cal{E}}(N,V_2)$, and $\mbox{vol}_{box}[{\cal{M}}(E,N,t\gg
t_0)]$ is the box-counting volume of ${\cal{M}}(E,N,t\gg t_0)$ which is the
same as the volume of $\overline{\cal{M}}$, see below.

To obtain $\mbox{vol}_{box}[{\cal{M}}(E,N,t\gg t_0)]$ we cover the $d$-dim.
sub-manifold ${\cal{M}}(t)$, here with $d=(6N-1)$, of the phase space by a
grid with spacing $\delta$ and count the number $N_\delta\propto
\delta^{-d}$ of boxes of size $\delta^{6N}$, which contain points of ${\cal
{M}}$.  Then we determine
\begin{eqnarray}
\mbox{vol}_{box}[{\cal{M}}(E,N,t\gg t_0)]&:=&\underbar{$\lim$}_{\delta\to
0} \delta^d N_\delta[{\cal{M}}(E,N,t\gg t_0)]\label{boxvol}\\
\lefteqn{\mbox{with }\underbar{$\lim *$}=\inf[\lim *]\mbox{ or
symbolically:}} \nonumber\\ M(E,N,t\gg t_0)&=:&
\displaystyle{B_d\hspace{-0.5 cm}\int}_{\{q_0\{q_t,p_t\}\subset V_1\}}
{\frac{1}{N!}\left(\frac{d^3q_t\;
d^3p_t}{(2\pi\hbar)^3}\right)^N\epsilon_0\delta(E-\hat H_N)}\label{boxM}\\
&\to&\frac{1}{N!}\int_{\{q_t\subset V_2\}} {\left(\frac{d^3q_t\;d^3p_t}
{(2\pi\hbar)^3}\right)^N\epsilon_0\delta(E-\hat H_N\{q_t,p_t\})}
\nonumber\\ &=&W(E,N,V_2) \ge W(E,N,V_1),
\end{eqnarray}
where $\displaystyle{B_d\hspace{-0.5 cm}\int}$ means that this integral
should be evaluated via the box-counting volume (\ref{boxvol}) here with
$d=6N-1$.

This is illustrated by the following figure (\ref{box}).
\begin{figure}
\begin{minipage}[t]{6cm}
\begin{center}$V_a$\hspace{2cm}$V_b$\end{center}
\vspace*{0.2cm}
\includegraphics*[bb = 0 0 404 404, angle=-0, width=5.7cm,
clip=true]{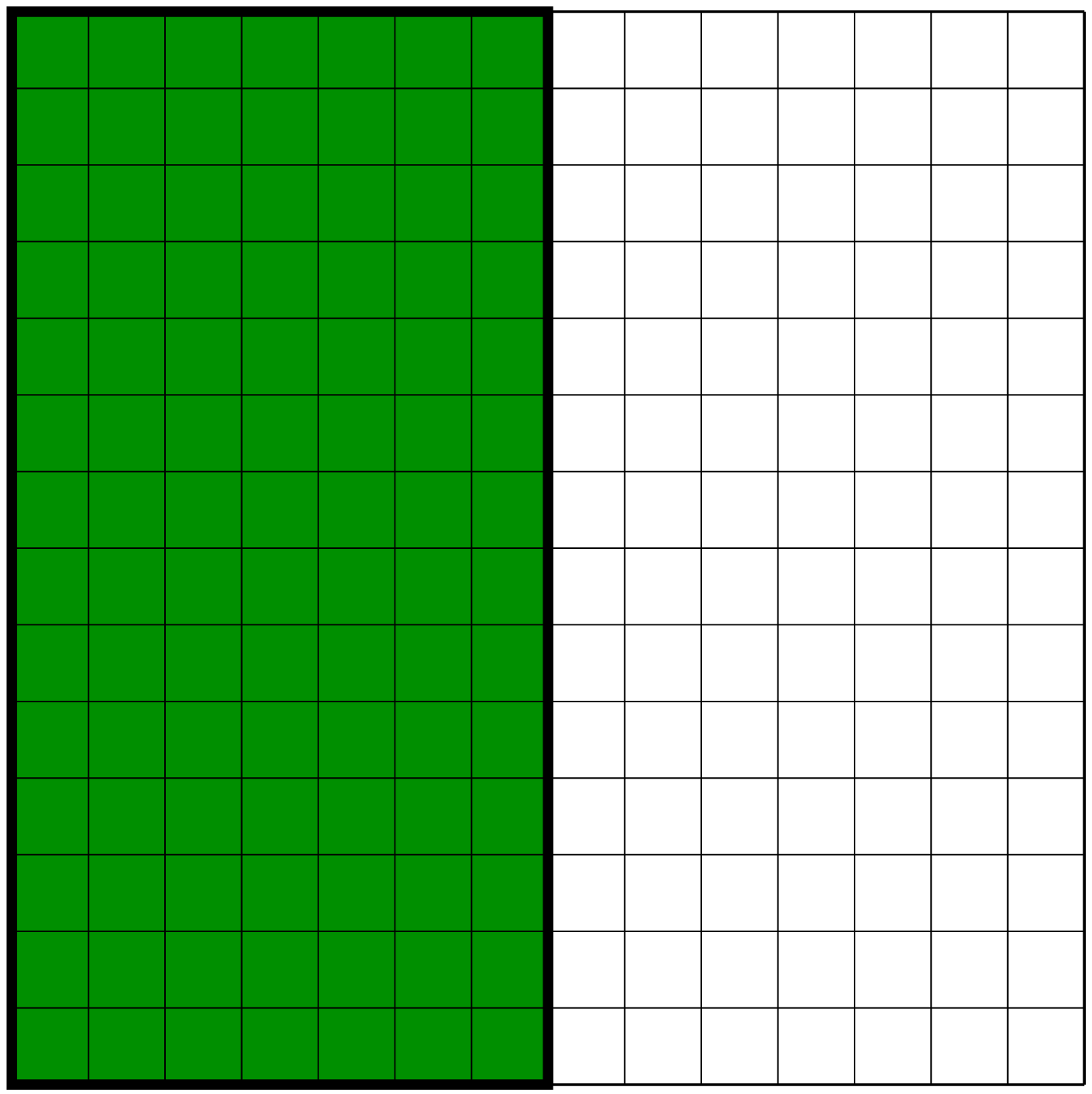}\begin{center}$t<t_0$\end{center}
\end{minipage}\lora\begin{minipage}[t]{6cm}
\begin{center}$V_a+V_b$\end{center}
\includegraphics*[bb = 0 0 428 428, angle=-0, width=6cm,
clip=true]{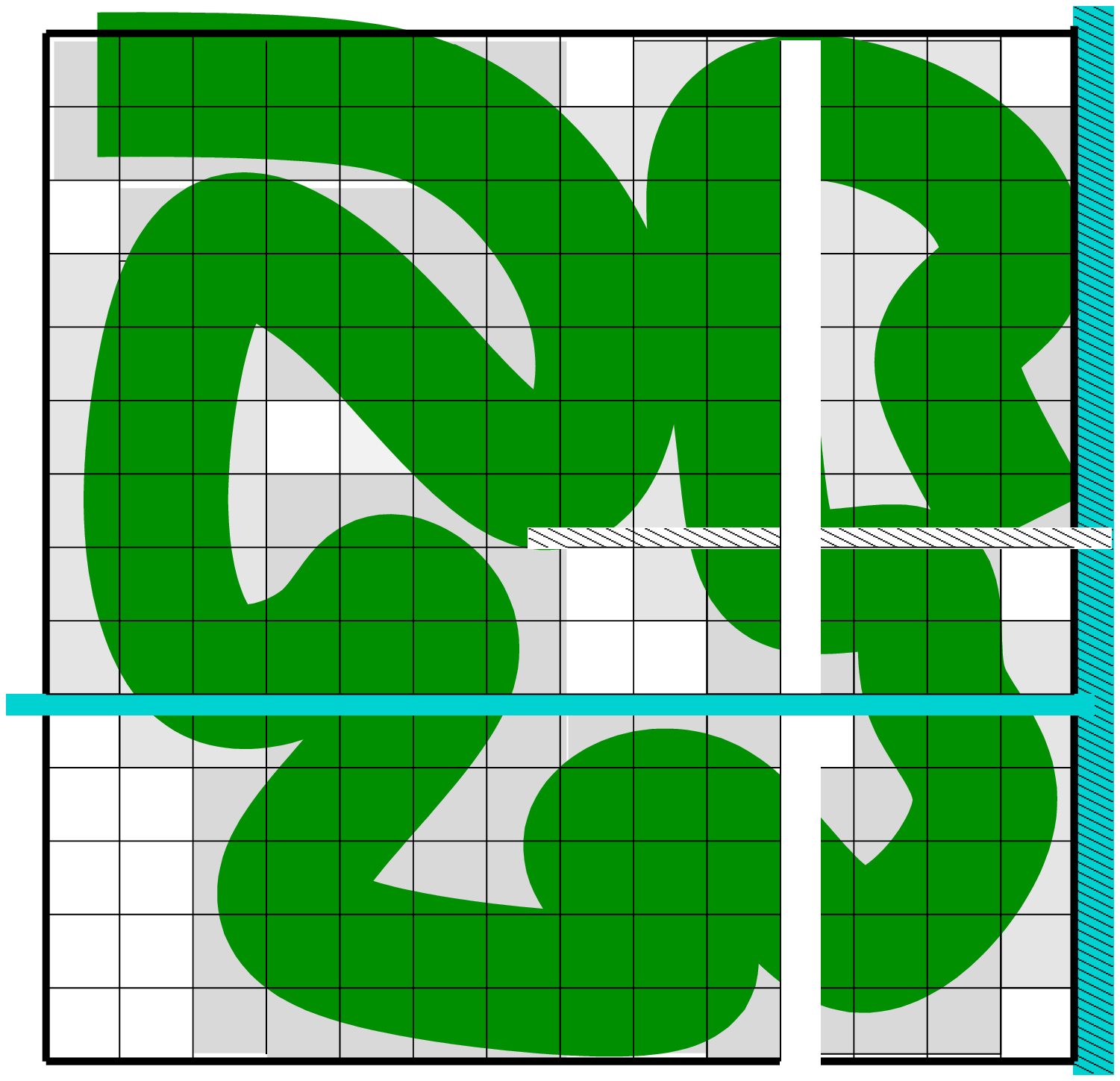}\begin{center}$t>t_0$\end{center}
\end{minipage}
\caption{ The compact set ${\cal{M}}(t_0)$, left side, develops
  into an increasingly folded ``spaghetti''-like distribution in
  phase-space with rising time $t$,  Gibbs' "ink-lines". The right figure shows only the
  early form of the distribution. At much larger times it will become
  more and more fractal and finally dense in the new phase space.  The
  grid illustrates the boxes of the box-counting method. All boxes
  which overlap with ${\cal{M}}(t)$ are counted in $N_\delta$ in
  eq.(\ref{boxvol})\label{box}}
\end{figure}
With this extension of eq.(\ref{phasespintegr}) Boltzmann's entropy
(\ref{boltzmentr1}) is at time $t\gg t_0$ equal to the logarithm of the
{\em larger} phase space $W(E,N,V_2)$. {\em This is the second law
  of thermodynamics.}  The box-counting is also used in the definition
of the Kolmogorov entropy, the average rate of entropy gain
\cite{falconer90,crc99}.  Of course still at $t_0$
$\overline{{\cal{M}}(t_0)}={\cal{M}}(t_0)={\cal{E}}(N,V_1)$:
\begin{eqnarray}
M(E,N,t_0) &=:&\displaystyle{B_d\hspace{-0.5 cm}\int}_{\{q_0\subset V_1\}}
{\frac{1}{N!}\left(\frac{d^3q_0\;
d^3p_0}{(2\pi\hbar)^3}\right)^N\epsilon_0\delta(E-\hat H_N)}\\
&\equiv&\int_{\{q_0\subset V_1\}} {\frac{1}{N!}\left(\frac{d^3q_0\;
d^3p_0}{(2\pi\hbar)^3}\right)^N\epsilon_0\delta(E-\hat H_N)}\nonumber\\
&=&W(E,N,V_1).
\end{eqnarray}

The box-counting volume is analogous to the standard method to determine
the fractal dimension of a set of points \cite{falconer90} by the
box-counting dimension:
\begin{equation}
\dim_{box}[{\cal{M}}(E,N,t\gg t_0)]:=\underbar{$\lim$}_{\delta\to 0}
\frac{\ln{N_\delta[{\cal{M}}(E,N,t\gg t_0)]}}{-\ln{\delta}}\\
\end{equation}

Like the box-counting dimension, $\mbox{vol}_{box}$ has the peculiarity
that it is equal to the volume of the smallest {\em
  closed} covering set. E.g.: The box-counting volume of the set of
rational numbers $\{{\bf Q}\}$ between $0$ and $1$, is
$\mbox{vol}_{box}\{{\bf Q}\}=1$, and thus equal to the measure of the {\em
real} numbers , c.f. Falconer \cite{falconer90} section 3.1. This is the
reason why $\mbox{vol}_{box}$ is not a measure in its strict mathematical
sense because then we should have
\begin{equation}
\mbox{vol}_{box}\left[\sum_{i\subset\{\bf Q\}}({\cal{M}}_i)\right]=
\sum_{i\subset\{\bf Q\}}\mbox{vol}_{box}[{\cal{M}}_i]=0,
\end{equation}
therefore the quotation marks for the box-counting ``measure'' c.f.
appendix \ref{app}.

Coming back to the the end of section (\ref{EPSformulation}), the volume
$W(A,B,\cdots,t)$ of the relevant ensemble, the {\em closure}
$\overline{{\cal{M}}(t)}$ must be ``measured'' by something like the
box-counting ``measure'' (\ref{boxvol},\ref{boxM}) with the box-counting
integral $\displaystyle{ B_d\hspace{-0.5 cm}\int}$, which must replace the
integral in eq.(\ref{phasespintegr}). Due to the fact that the box-counting
volume is equal to the volume of the smallest closed covering set, the new,
extended, definition of the phase-space integral eq.(\ref{boxM}) is for
compact sets like the equilibrium distribution ${\cal{E}}$ identical to the
old one eq.(\ref{phasespintegr}) and nothing changes for equilibrium
statistics. Therefore, one can simply replace the old Boltzmann-definition
of the number of complexions and with it of the entropy by the new one
(\ref{boxM}).

\subsection{Conclusion of the discussion of the second law}
Macroscopic measurements $\hat{M}$ determine only a very few of all $6N$
d.o.f.  Any macroscopic theory like thermodynamics deals with the {\em
volumes} $M$ of the corresponding closed sub-manifolds $\overline{\cal{M}}$
in the $6N$-dim. phase space not with single points.  The averaging over
ensembles or finite sub-manifolds in phase space becomes especially
important for the microcanonical ensemble of a {\em finite} system.

Because of this necessarily coarsed and redundant information, macroscopic
measurements, and with it also macroscopic theories are unable to
distinguish fractal sets ${\cal{M}}$ from their closures
$\overline{\cal{M}}$. Therefore, I make the conjecture: the proper
manifolds determined by a macroscopic theory like thermodynamics are the
closed $\overline{\cal{M}}$. However, an initially closed subset of points
at time $t_0$ does not necessarily evolve again into a closed subset at
$t\gg t_0$. I.e. the closure operation and the $t\to\infty$ limit do not
commute, and the macroscopic dynamics becomes thus irreversible.

Here is the origin of the misunderstanding by the famous reversibility
paradoxes which were invented by Loschmidt \cite{loschmidt76} and Zermelo
\cite{zermelo96,zermelo97} and which bothered Boltzmann so much
\cite{cohen97,cohen00}. These paradoxes address to trajectories of {\em
single points} in the $N$-body phase space which must return after
Poincarre's recurrence time or which must run backwards if all momenta are
exactly reversed. Therefore, Loschmidt and Zermelo concluded that the
entropy should decrease as well as it was increasing before. The
specification of a single point demands of course a {\em microscopic exact}
specification of all $6N$ degrees of freedom not a determination of a few
macroscopic degrees of freedom only. No entropy is defined for a single
point. Moreover, it is highly unlikely that all points in the
micro-ensemble have commensurate recurrence times so that they can return
{\em simultaneously} to their initial positions. Once the {\em manifold}
has spread over the larger phase space it will never return.

This way various non-trivial limiting processes are avoided. Neither does
one invoke the thermodynamic limit of a homogeneous system with infinitely
many particles nor does one rely on the ergodic hypothesis of the
equivalence of (very long) time averages and ensemble averages.  {\em The
use of ensemble averages is justified
  directly by the very nature of macroscopic (incomplete)
  measurements}. Coarse-graining appears as natural consequence of
this. The box-counting method mirrors the averaging over the overwhelming
number of non-determined degrees of freedom. Of course, a fully consistent
theory must use this averaging explicitly. Then one would not depend on the
order of the limits $\lim_{\delta\to
  0}\lim_{t\to\infty}$ as it was tacitly assumed here. Presumably, the
rise of the entropy can then be already seen at finite times when the
fractality of the distribution in phase space is not yet fully developed.
The coarse-graining is no more a mathematical ad hoc assumption.  Moreover
the second law is in the EPS-formulation of statistical mechanics not
linked to the thermodynamic limit as was thought up to now
\cite{lebowitz99,lebowitz99a,goldstein03}.

\section{Final conclusion}
Entropy has a simple and elementary definition by the {\em area}
$e^{S(E,N,\cdots)}$ of the microcanonical ensemble in the $6N$ dim. phase
space. Canonical ensembles are not equivalent to the micro-ensemble in the
most interesting situations:
\begin{enumerate}
\item at phase-separation (\lora heat engines !), one gets
 inhomgeneities, and a negative heat capacity or some other negative susceptibility,
\item  Heat can flow from cold to hot.
\item phase transitions can be localized sharply and unambiguously in
small classical or quantum systems, there is no need for finite size
scaling to identify the transition.
\item also really large self-gravitating systems can now be addressed.
\end{enumerate}
Entropy rises during the approach to equilibrium, $\Delta S\ge 0$, also for
small mixing systems. i.e. the second law is valid even for small systems
\cite{gross183,gross192}.

With this geometric foundation thermo-statistics applies not only to
extensive systems but also to non-extensive ones which have no
thermodynamic limit.
\section{Appendix}\label{app}
In the mathematical theory of fractals\cite{falconer90} one usually uses
the Hausdorff measure or the Hausdorff dimension of the fractal
\cite{crc99}.  This, however, would be wrong in Statistical Mechanics. Here
I want to point out the difference between the box-counting ``measure'' and
the proper  Hausdorff measure of a manifold of points in phase space.
Without going into too much mathematical details we can make this clear
again with the same example as above: The Hausdorff measure of the rational
numbers $\in[0,1]$ is $0$, whereas the Hausdorff measure of the real
numbers $\in[0,1]$ is $1$. Therefore, the Hausdorff measure of a set is a
proper measure. The Hausdorff measure of the fractal distribution in phase
space ${\cal{M}}(t\to\infty)$ is the same as that of ${\cal{M}}(t_0)$,
$W(E,N,V_1)$. Measured by the Hausdorff measure the phase space volume of
the fractal distribution ${\cal{M}}(t\to\infty)$ is conserved and
Liouville's theorem applies. This would demand that thermodynamics could
distinguish between any point inside the fractal from any point outside of
it independently how close it is.  This, however, is impossible for any
macroscopic theory that can only address the redundant macroscopic
information where all unobserved degrees of freedom are averaged over. That
is the deep reason why the box-counting ``measure'' must be taken and where
irreversibility comes from.




%
\nocite{*}
\bibliographystyle{plain}
\bibliography{Surface}
\def\footmsgA{{\copyright 2004 by MDPI (http://www.mdpi.org). Reproduction for noncommercial purposes permitted.}}
\end{document}